\documentclass[journal]{IEEEtran}
\usepackage{color}
\usepackage{graphicx}
\usepackage{epstopdf}
\usepackage{amssymb,amsmath}
\usepackage{amsthm}
\usepackage{cite}
\usepackage{multirow}
\usepackage{relsize}
\usepackage{subfigure}
\usepackage[hidelinks]{hyperref}
\usepackage[rgb]{xcolor}
\usepackage[author={R. Abbas}]{pdfcomment}
\bstctlcite{IEEEexample:BSTcontrol}
\newtheorem{proposition}{Proposition}

\newtheorem{lemma}{Lemma}

\theoremstyle{definition}

\newcommand\blfootnote[1]{%
  \begingroup
  \renewcommand\thefootnote{}\footnote{#1}%
  \addtocounter{footnote}{-1}%
  \endgroup
}
\IEEEaftertitletext{\vspace{-2\baselineskip}}

\ifCLASSINFOpdf

\else

\fi

\begin{document}

\title{Random Multiple Access for M2M Communications with QoS Guarantees}

\author{Rana~Abbas,~\IEEEmembership{Student Member,~IEEE,}~Mahyar~Shirvanimoghaddam,~\IEEEmembership{Member,~IEEE,}~Yonghui~Li,~\IEEEmembership{Senior Member,~IEEE,}~and~Branka~Vucetic,~\IEEEmembership{Fellow,~IEEE}}
\maketitle
\begin{abstract}
We propose a novel random multiple access (RMA) scheme with quality of service (QoS) guarantees for machine-to-machine (M2M) communications. We consider a slotted uncoordinated data transmission period during which machine type communication (MTC) devices transmit over the same radio channel.
{Based on the latency requirements, MTC devices are divided into groups of different sizes, and the transmission frame is divided into subframes of different lengths.} In each subframe, each group is assigned an access probability based on which an MTC device decides to transmit replicas of its packet or remain silent. The base station (BS) employs successive interference cancellation (SIC) to recover all the superposed packets.
We derive the closed form expressions for the average probability of device resolution for each group, and we use these expressions to design the access probabilities. The accuracy of the expressions is validated through Monte Carlo simulations. We show that the designed access probabilities can guarantee the QoS requirements with high reliability and high energy efficiency. Finally, we show that RMA can outperform standard coordinated access schemes as well as some of the recently proposed M2M access schemes for cellular networks.
\end{abstract}
\section{Introduction}
\blfootnote{R. Abbas, Y. Li and B. Vucetic are with the Center of Excellence in Telecommunications, School of Electrical and Information Engineering, The University of Sydney, NSW, 2006, Australia (email: \{rana.abbas;yonghui.li;branka.vucetic\}@sydney.edu.au).}
\blfootnote{M. Shirvanimoghaddam is with the School of Electrical Engineering and Computer Science, The University of Newcastle, NSW, 2308, Australia (email:mahyar.shirvanimoghaddam@newcastle.edu.au).}
\blfootnote{This work was partly presented at ISIT 2016.}
\IEEEPARstart{M}{achine}-to-machine (M2M) communications involve the communication between machine type communication (MTC) devices in a fully automated fashion without or with little human intervention. M2M communications have a wide range of applications ranging from smart metering, intelligent transportation systems, health monitoring to energy management and smart grids, etc. Unlike human-to-human (H2H) communications which involve a relatively small number of devices with high data rate requirements and large message size, M2M communications are generally characterized by a massive number of devices with low-data rate sporadic transmissions of small payloads. Moreover, most MTC devices have fixed locations with low computational and storage capabilities as well as a low power budget. Although some of these requirements have been considered in the current wireless standards, the unique requirements and traffic characteristics of M2M require major changes in the air interface and core network. This has been identified as a key area in which major changes are required to enable cellular systems to handle M2M traffic \cite{survey1}.
\subsection{Related Work}
The third generation partnership project (3GPP) has identified issues and challenges related to M2M communications in the future release of Long Term Evolution (LTE), referred to as LTE-Advanced (LTE-A) \cite{3GPPLTEA}. In LTE-A, devices are required to establish an air interface connection prior to data transmission. Access requests are transmitted in an uncoordinated manner over the random access channel (RACH). Once an MTC device has been granted access, it is scheduled to specific radio resources over which data transmission takes place in a deterministic manner. However, with the massive number of MTC devices forecasted to operate in the near future, the RACH in current access schemes will be overloaded and will suffer from continuous collisions requiring multiple re-transmissions. This will result in a large energy expenditure, unexpected delays, and time-frequency resource wastage. Some solutions have been proposed and even standardized to mitigate the RACH overload problem \cite{survey1,survey2} e.g. access class barring schemes, separate RACH resources for M2M, dynamic allocation of RACH resources, backoff schemes, slotted access, pull-based and group-based schemes.

Aside from the RACH overload problem, establishing energy efficient access schemes with QoS guarantees is another design challenge in M2M networks. We list some of the works proposed in this direction. A group-based approach was proposed in \cite{lien2011toward} where devices are grouped based on their latency requirements. Each group is allocated specific time intervals with predefined duration over which data transmission takes place in a deterministic manner. The allocated time intervals are non-overlapping and their periodicity is proportional to their packet arrival rate. Thus, groups with higher packet arrival rates are assumed to have tighter latency requirements. An extension of this work can be found in \cite{si2014adaptive} where this assumption is dropped and a generalized access management scheme was proposed with an adaptive resource allocation scheme based on the incoming traffic. In \cite{shirvanimoghaddam2015probabilistic}, devices with the same QoS requirement are scheduled to transmit over the same resources simultaneously. Using well-designed codes, the superposed codewords can be reliably decoded at the base station (BS) allowing for a more efficient use of resources and reduction in the number of retransmissions at the RACH.
Other approaches \cite{Dhillon,zhang2014tree,kahn2015connectionless} have studied uncoordinated access schemes which do not use any RACHs and instead assign all the resources as uplink data channels. These works have shown potential performance gains in supporting a larger number of devices in comparison to coordinated access schemes for small packet transmissions. In the following, we review the recent works tackling the massive access problem through uncoordinated access which is the  main focus of this paper.

For a single cell scenario with a large number of devices contending to access the same BS, a slotted protocol has been widely studied \cite{roberts1975aloha}. An uncoordinated data transmission period is initiated by the BS where each device transmits multiple copies of its packet in different time slots chosen independently and uniformly at random. The BS stores all the observed slots and sorts them into three categories: idle slots (no packets received), singleton slots (one packet received) and collision slots (multiple packets received). In conventional slotted ALOHA, packets can only be recovered if an interference-free copy has been received at the BS, i.e., singleton slots. Therefore, idle slots are wasteful of resources, and collision slots are wasteful of energy. However, by employing successive-interference cancellation (SIC), one can cancel the recovered packets from the collision slots allowing for the resolution of a larger fraction of devices \cite{sic,sigsag,IRSA}. In \cite{IRSA}, the number of transmissions of a given device is a random variable whose distribution is predetermined at the BS and broadcasted at the beginning of each transmission period. Authors draw an analogy between this scheme and the decoding of codes-on-graph for binary erasure channels \cite{lt}. Thus, the same analytical tools, namely the AND-OR tree \cite{andor}, can be used to find the necessary probabilities and the maximum load that can be supported. Many extensions have been proposed since that have further emphasized the scalability of RMA with SIC and its capability of supporting a large number of devices without prior identification \cite{popovskiLetter,popovskiRateless,popovskiWCNC2014,abbas2015,abbasGC2015}. By requiring little signalling overhead and centralized processing, it is being considered as a possible candidate for future M2M communications \cite{zorziGC}.

{Authors in \cite{stefanovic2013coded} extended this model to a heterogenous slotted ALOHA setting using the extended AND-OR tree framework derived in \cite{sejdinovic2010decentralised}. In \cite{stefanovic2013coded}, the heterogeneity in the network represented the different packet loss rates amongst the devices corresponding to their different channel conditions. Using a similar analytical framework, authors in \cite{toni2015prioritized} considered the heterogenous QoS requirements in the network and formulated an optimization problem to find the probabilities and the system load that maximize the overall system utility. All these schemes can be seen as a special case of our more generalized framework which is shown to be more suitable for M2M applications. We extend the analytical expressions and reformulate the designs and optimization problems to satisfy the unique requirements and design constraints of M2M communications.}
\subsection{Contributions and Organization}
The main contributions of this paper are summarized in what follows.
\subsubsection{Random Multiple Acess Scheme}
We propose a generalized slotted uncoordinated data transmission scheme for the case of diverse QoS requirements. We consider two different schemes where devices are grouped based on their QoS requirements.
{The first scheme is called the ACK-All scheme. In this scheme, MTC devices from all groups transmit simultaneously over the same radio resources in all stages of the transmission frame. The second scheme is called the ACK-Group scheme. In this scheme, MTC devices from the same group transmit in distinct stages.} We discuss the advantages of these schemes over coordinated access schemes. We show that the proposed RMA schemes can service a larger number of devices over the same number of resources when the number of resources is sufficiently large, while guaranteeing the diverse QoS requirements. We further show that the ACK-All scheme is advantageous over the ACK-Group scheme when the  number of devices with tighter latency requirements are less than those with more flexible latency requirements.
\subsubsection{AND-OR Tree Based Performance Analysis}
{We use the AND-OR tree to analyze the system performance expressed by the average probability of device resolution error for both transmission schemes: ACK-All and ACK-Group. These expressions can be seen as a generalization of those in [24] which only characterizes the error probabilities at the end of a single subframe. As the devices that are acknowledged do not retransmit in following subframes, the intermediate feedback introduced between the subframes results in graphs with reduced sets of nodes as well as reduced edges. We characterize these reductions by reformulating the AND-OR tree expressions based on the reduced sets of active devices and slots along with their reduced degree distributions. We validate the accuracy of the expressions under different settings using simulations.}
\subsubsection{{System Design for QoS Guarantees}}
{We show how the derived expressions can be used to design systems that can guarantee the QoS requirements of different groups with significantly high energy efficiency and high reliability for the proposed scheme. For this, we cannot use the optimization problem formulated in \cite{toni2015prioritized}. Authors in \cite{toni2015prioritized} target vanishing error probabilities. However, the error probabilities cannot be assumed to be vanishing at the end of each subframe in the proposed scheme. For example, lower priority groups can tolerate larger delays and, thus, their error probabilities will be far from vanishing in the early stages of transmission. Moreover, vanishing error probabilities are only relevant to ultra-reliable applications which are only a subset of M2M applications. Therefore, we need to guarantee the diverse latency requirements with diverse error probabilities. Accordingly, we propose a guideline that allows us to design the access probabilities using the AND-OR tree, which assumes asymptotically large systems, for a finite number of devices and resources.}

{The rest of this paper is organized as follows. In Section II, we present the system model and the proposed RMA schemes. In Section III,  we consider a tree-based analytical framework and derive the expressions for the average probabilities of device resolution. An energy efficient system design is discussed in Section IV along with the limitations on the system load. A reliable system design is discussed in Section V and is shown to be valid for a finite number of devices. Numerical results and practical considerations are presented in Section VI. Conclusions are drawn in Section VII.}
\section{System Model}
\subsection{Overview}
We consider a scenario where $K$ uniformly distributed MTC devices communicate with a BS located at the origin. The devices are assumed to have fixed locations. We consider the uplink of an orthogonal frequency division multiple access (OFDMA) system, where the frequency resources are divided into several sub-channels each with a bandwidth $\Delta f$. A radio resource unit consists of a sub-channel along with a time slot of duration $T$. We characterize the QoS of MTC devices in terms of the delay as diverse M2M applications have diverse latency requirements. More specifically, we consider different delay groups denoted by $\mathcal{C}_1, \mathcal{C}_2,...,\mathcal{C}_r$, and their respective delay requirements are quantified in terms of time slots and denoted by $N_1,N_2,\ldots,N_r$, where $N_i< N_{i'}$ for $i < i'$, i.e., devices from the group $\mathcal{C}_i$ have a tighter latency requirement than devices from the group $\mathcal{C}_{i'}$ for $i<i'$. The notations used in this paper are summarized in Table \ref{not} for quick reference.

The resources allocated for M2M may either be fixed or may vary from one frame to the other. For every transmission frame, the BS is assumed to be capable of estimating the number of active devices and grouping them based on their QoS requirements. Similarly, the number of active devices and the number of groups may either be fixed or may vary from one frame to the other. Active devices are devices with at least one packet to transmit. Furthermore, we assume devices transmit one packet within one transmission frame. Packets are assumed to be of equal size and can be transmitted in one radio resource unit.

We assume a block fading channel, i.e., the channel gains remain unchanged for the duration of the transmission frame and vary randomly and independently from one frame to the other. Channels are also assumed to be reciprocal, i.e., uplink and downlink channel states are the same. Therefore, each device can estimate the uplink channel gain from the pilot signal sent periodically over the downlink channel by the BS. The devices also use these pilot signals to synchronize their timing to that of the BS \cite{wang2011green,chen2014time}. The devices perform channel inversion such that they all have the same received signal-to-noise ratio (SNR) $\gamma_{\mathrm{ref}}$ \cite{si2014adaptive}. $\gamma_{\mathrm{ref}}$ is determined by the coding and modulation schemes adopted in the system such that the BS can reliably recover a device's packet if its SNR is greater than or equal to $\gamma_{\mathrm{ref}}$. In reality, MTC devices are power limited. For channels with low SNR, the device may not always be able to perform channel inversion. In this case, the device's channel is said to be in outage and the device remains silent.
\subsection{Probabilistic Data Transmission}
\begin{table}[!t]
  \centering
  \caption{Notation Summary}
  \begin{tabular}{|p{0.9cm}| p{7cm}|}
  \hline
  \textbf{Notation}&\textbf{Description}\\
  \hline
  $K$ &Number of MTC devices\\
  \hline
  $r$& Number of groups of MTC devices\\
  \hline
  $N_i$&Latency requirement of the $i^{th}$ group of MTC Devices\\
  \hline
  $\mathcal{C}_i$&Group of MTC devices with a latency requirement of $N_i$ \\
  \hline
  $\mathcal{C}_{i,k}^{(s)}$& Subset of MTC devices in $\mathcal{C}_i$ which transmitted in $k$ subframes, $k\leq s$\\
  \hline
  $\Delta N_i$&Number of times slots in subframe $i$\\
  \hline
  $N$ &Total number of time slots in a frame\\
  \hline
  $p_i^{(s)}$ &Access probability allocated to $\mathcal{C}_i$ in subframe $s$\\
  \hline
  $g_i^{(s)}$&Average degree of a check node in subframe $s$\\
  \hline
  $\zeta_i^{(s)}$&Average degree of a variable node in $\mathcal{C}_{i,s}^{(s)}$\\
  \hline
  $\epsilon_i^{(s)}$&Average probability of device resolution error of $\mathcal{C}_i$ after $s$ subframes\\
  \hline
  $\epsilon_i^*$&Target probability of device resolution error of $\mathcal{C}_i$\\
  \hline
  $\alpha_i$&Fraction of MTC devices in $\mathcal{C}_i$\\
  \hline
  $\beta_s$&Fraction of time slots in subframe $s$\\
  \hline
  $M_i$&Average number of transmissions of a device in $\mathcal{C}_i$\\
  \hline
  $\gamma_{\mathrm{ref}}$&Received SNR of a device's packet\\
  \hline
  \end{tabular}\label{not}
\end{table}

Each transmission frame is of length $N = N_r$ time slots. The frame is divided into $r$ subframes, where the length of subframe $s$ is denoted by $\Delta N_s $ and is equal to $N_s - N_{s-1}$, for $1\leq s \leq r$ and $N_0 = 0$. For a given subframe $s$, the BS assigns every group $\mathcal{C}_i$ an access probability $p_{i}^{(s)}$. That is, a device in group $\mathcal{C}_i$ transmits in a given time slot of subframe $s$ with a probability $p_i^{(s)}$ and remains silent with a probability $1-p_i^{(s)}$.
{We emphasize that the number of probabilities that are to be distributed to the devices is equal to the number of different groups and not the number of devices. Therefore, the signaling overhead is fairly small and can be easily incorporated alongside the pilot signals, beacons, acknowledgements and other control data.}

The BS is assumed to be able to distinguish between idle slots, singleton slots, and collision slots. This is feasible with the assumed power control strategy. Once the BS detects the reception of a singleton slot, the transmitted packet is recovered and the respective device is said to be resolved. The BS and the devices can share a common pseudo-random generator function that computes a random seed based on the device's identity. The devices use these seeds to generate the indices of the slots they want to transmit in. Once a device's packet is recovered at the BS, the BS can extract the respective device's identity from the header file. Then, the BS can generate the seed and the necessary indices of slots to perform SIC. That is, the copies of the recovered packet are cancelled from the remaining slots. This allows for the potential of having more singleton slots in the following iteration, and, thus, the potential of recovering more packets.

\begin{figure}[!t]
  \centering
  \includegraphics[width=3.00in]{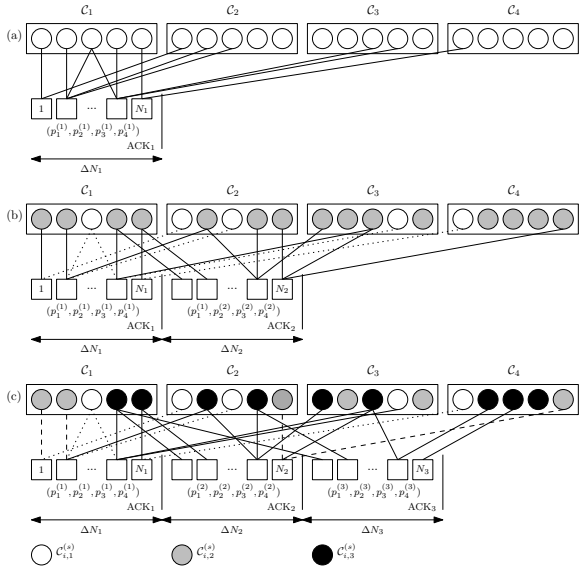}\\
  \caption{Bipartite graphs of the ACK-All scheme with $r = 4$ in three consecutive subframes. The dotted and dashed lines denote the edges that were removed at the receiver side after the first and second subframe respectively, and the solid lines denote the edges that have remained.}\label{BP2}
\end{figure}
The BS performs SIC at the end of each subframe $s$ using all the received packets over the $N_s$ time slots. We assume a perfect feedback channel. A batch of acknowledgements is sent at the end of each subframe to all successfully resolved devices, and all acknowledged devices do not transmit in future time slots. We refer to this scheme as the ACK-All scheme.
{The transmission scheme in \cite{toni2015prioritized} can be seen as a special case with a single subframe of our more generalized ACK-All scheme. The main motivation behind dividing the transmission frame into subframes separated by intermediate feedback is to better service the lower prioritized devices which suffer from a degraded performance when sharing resources with more prioritized devices. This behaviour has been noted for codes-on-graph \cite{UEP_Rahnavard} as well as random access \cite{toni2015prioritized}.}

An example of the ACK-All scheme is illustrated in Fig. \ref{BP2} in terms of a bipartite graph. The circles represent the devices and are referred to as variable nodes (VNs). The squares represent the time slots and are referred to as check nodes (CNs). An edge between two nodes indicates that the device transmitted in the respective time slot. No edge between two nodes indicates that the device was silent in the respective time slot. As mentioned before, the BS performs SIC by cancelling the copies of the recovered packet from the remaining slots. In that case, the edges connecting the resolved VN to the CNs are removed from the graph. Moreover, we denote by $\mathcal{C}_{i,k}^{(s)}$ the subset of devices in group $\mathcal{C}_i$ that transmitted in the first $k$ subframes, where $1\leq k \leq s$.

Fig. \ref{BP2} assumes flexible latency requirements as unresolved devices continue to transmit even after their delay has been violated. For stricter latency requirements, unresolved devices are dropped from the network at that point. This is illustrated in Fig. \ref{BP_compare}(a). In Fig. \ref{BP_compare}(b), we consider a variant of the ACK-All scheme: the case of separate transmissions. We refer to this scheme as the ACK-Group scheme where devices from the group $\mathcal{C}_i$ are scheduled to transmit in subframe $i$ only and remain silent otherwise. Thus, transmissions of devices from different groups are shown to be separated.
\section{AND-OR Analysis of the Proposed RMA Schemes}
The number of edges branching out of a node is said to be the degree of the respective node. These degrees are random variables that play an important role in the system performance. In this section, we first derive the expressions of the degree distributions of the VNs and the CNs. We also introduce the AND-OR tree which is a well-known tool for calculating the decoding error probabilities of information symbols in codes on graph. Then, we propose a more generalized AND-OR tree to model the system with the ACK-All and ACK-Group schemes.
\subsection{Degree Distributions}
\begin{figure}[!t]
  \centering
  \includegraphics[width=3.0in]{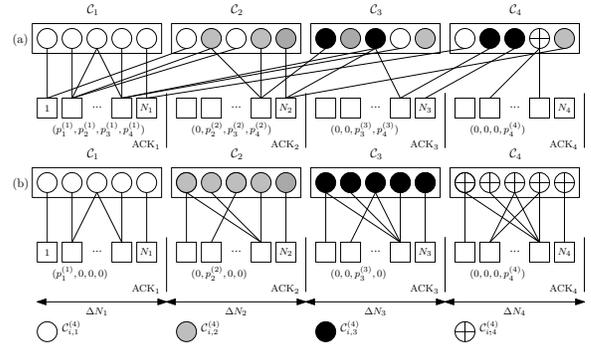}\\
  \caption{Overall bipartite graph representation of the two transmission schemes with $r = 4$: (a) ACK-All and (b) ACK-Group.}\label{BP_compare}
\end{figure}
For generality purposes, we assume flexible latency requirements, i.e., $p_i^{(s)} \geq 0$ for $i>s$. We denote by $\mathcal{G}_s$ the bipartite graph formed by the first $s$ subframes. As mentioned before, for a given subframe $s$, each group $\mathcal{C}_i$ can be divided into $s$ subsets: $\mathcal{C}_{i,1}^{(s)},\mathcal{C}_{i,2}^{(s)},\ldots,\mathcal{C}_{i,s}^{(s)}$.

The probability that a device from $\mathcal{C}_{i,s}^{(s)}$ transmits in a given time slot of subframe $s$ is a Bernoulli random variable with a success probability of $p_i^{(s)}$. As this probability is independent of previous transmissions, the probability that a device from the group $\mathcal{C}_{i,s}^{(s)}$ transmits in $d$ time slots in a subframe $s$ is the sum of $d$ i.i.d. Bernoulli random variables and is denoted by $\Lambda_{i,d}^{(s)}$. This is equivalent to the following binomial distribution:
\begin{align}\label{VN_bino}
\Lambda_{i,d}^{(s)} = \begin{pmatrix}  \Delta N_s\\d\end{pmatrix}\left(p_i^{(s)}\right)^d\left(1-p_i^{(s)}\right)^{\Delta N_s-d}.
\end{align}
Furthermore, the probability that $d$ devices from the group $\mathcal{C}_{i,s}^{(s)}$  transmit in a given time slot of subframe $s$, is also the sum of $d$ i.i.d. Bernoulli random variables. The equivalent binomial distribution is given as
\begin{align}\label{CN_bino}
\Omega_{i,d}^{(s)} = \begin{pmatrix} |\mathcal{C}_{i,s}^{(s)}|\\d\end{pmatrix}\left(p_i^{(s)}\right)^d\left(1-p_i^{(s)}\right)^{|\mathcal{C}_i^{(s)}|-d}.
\end{align}

For convenience, let $p_i^{(s)} = \frac{g_i^{(s)}}{|\mathcal{C}_{i,s}^{(s)}|}$, where $0 \leq g_{i}^{(s)}\leq |\mathcal{C}_{i,s}^{(s)}|$ is the average number of devices from the group $\mathcal{C}_{i,s}^{(s)}$ that access a given time slot of subframe $s$. For sufficiently large $K$ and $N$, the expression in (\ref{VN_bino}) can be approximated by the following Poisson distribution
\begin{equation}\label{VN_poiss}
\Lambda_{i,d}^{(s)}= \frac{{\left(\zeta_i^{(s)}\right)}^d\exp{\left(-\zeta_i^{(s)}\right)}}{d!},\quad \text{for }1\leq k\leq s,
\end{equation}
where $\zeta_i^{(s)} = \frac{g_i^{(s)}\Delta N_{s}}{|\mathcal{C}_{i,s}^{(s)}|}$ is the average number of time slots from subframe $s$ that are selected by a given device from the group $\mathcal{C}_{i,s}^{(s)}$. Similarly, the expression in (\ref{CN_bino}) can be approximated by the following
\begin{equation}\label{CN_poiss}
\Omega_{i,d}^{(s)}= \frac{{\left(g_{i}^{(s)}\right)}^d\exp{\left(-g_i^{(s)}\right)}}{d!}.
\end{equation}

We now consider the generator polynomial $\Psi_{i}^{(s)}(x)=\sum_d \Psi_{i,d}^{(s)}x^d$ to represent the overall degree distribution of a VN in group $\mathcal{C}_{i,s}^{(s)}$ in the bipartite graph $\mathcal{G}_s$. With a slight abuse of notation, the superscript here denotes the sum degree over the first $s$ subframes rather than that of a single subframe $s$ as in (\ref{VN_bino}). For every VN, we define a vector $\textbf{m}$ of dimension $s$, where its $i^{th}$ element, $m_i$, corresponds to its number of transmissions in the $i^{th}$ subframe. $m_i$ is an independent binomial random variable with a distinct success probability as given in (\ref{VN_bino}). The overall degree of a VN is the sum of all elements of $\textbf{m}$ and, thus, follows a Poisson binomial distribution. We also define $\mathcal{M}_j^{(s)} \triangleq \{\textbf{m}:\sum_{k=1}^s m_k= j\}\text{, for }1\leq j\leq N_s$. The set $\mathcal{M}_j^{(s)}$ contains all possible realizations of $\textbf{m}$ whose elements add up to an overall degree of $j$. Therefore, in the subgraph $\mathcal{G}_s$, the probability that a VN of group $\mathcal{C}_{i,s}^{(s)}$ is of degree $j$ can be expressed as
\begin{align}\label{VN_prod}
\Psi_{i,j}^{(s)} = \sum_{\textbf{m}\in\mathcal{M}_j^{(s)}}\prod_{s' = 1}^{s}\Lambda_{i,m_{s'}}^{(s')}.
\end{align}

We also consider the generator polynomial $\Delta^{(s)}(x) = \sum_d \Delta_d^{(s)}x^d$ to represent the overall degree distribution of a CN in subframe $s$. For every CN, we define a vector $\textbf{l}$ of dimension $r$, where its $i^{th}$ element, ${l}_i$, corresponds to the number of transmissions from the $i^{th}$ group, $\mathcal{C}_i$. $l_i$ is an independent binomial random variable with a distinct success probability as given in (\ref{CN_bino}). Thus, the overall degree also follows a Poisson binomial distribution. We also define $\mathcal{L}_j^{(s)} \triangleq \{\textbf{l}:\sum_{i=1}^r l_i= j\}\text{, for }1\leq j\leq \sum_{i=1}^r |\mathcal{C}_{i,s}^{(s)}|$. $\mathcal{L}_j^{(s)}$ contains all possible realizations of $\textbf{l}$ whose elements add up to an overall degree of $j$. Then, we can write
\begin{equation}\label{CN_prod}
\Delta_j^{(s)} = \sum_{\textbf{l}\in\mathcal{L}_j^{(s)}}\prod_{i = 1 }^{r}\Omega_{i,l_i}^{(s)},\quad\text{for } 1\leq l\leq r.
\end{equation}

For sufficiently large $K$ and $N$, the sum degree can be seen as the sum of independent Poisson random variables with different averages, which is also a Poisson random variable with an average equal to the sum of its individual components. Accordingly, we can rewrite (\ref{VN_prod}) as
\begin{equation}\label{VN_poiss_all}
\Psi_{i}^{(s)}(x) = \exp\left(-\sum_{s'=1}^{s}\zeta_{i}^{(s')}(1-x)\right),
\end{equation}
Based on the same argument, we can rewrite (\ref{CN_prod}) as
\begin{equation}
\Delta^{(s)}(x) = \exp\left(-\sum_{i=1}^{r}g_{i}^{(s)}(1-x)\right).
\end{equation}
From an edge perspective, the generator polynomials corresponding to the degree distributions of both VNs and CNs are defined as follows:
\begin{equation}\label{edge_degree}
\psi_{i}^{(s)}(x) \triangleq \frac{\sum_d d\Psi_{i,d}^{(s)}x^{d-1}}{\bar{\Psi}_i^{(s)}}  \text{ and } \delta^{(s)}(x) \triangleq \frac{\sum_d d\Delta_{d}^{(s)} x^{d-1}}{\bar{\Delta}^{(s)}},
\end{equation}
where $\bar{\Psi}_i^{(s)} = \sum_d d\Psi_{i,d}^{(s)}$ and $\bar{\Delta}^{(s)} = \sum_d d\Delta_{d}^{(s)}$ are the average degrees of the devices and the slots in subframe $s$, respectively.
\subsection{Tree Assumption}
{Consider an edge $(v,c)$ chosen uniformly at random from $\mathcal{G}_s$ connecting a VN $v$ to an arbitrary CN $c$. By removing that edge, a subgraph is generated by $v$ and all the neighbors of $v$ within distance $2\ell$. This subgraph was shown in \cite{andor} to be a tree asymptotically with $v$ being its root at depth 0 and its leaves at depth $2\ell$.} Nodes at depth $i$ have children at depths $i+1$. The CNs are located at depth $1,3,...,2\ell-1$, and the VNs are located at depth $0,2,...,2\ell$. For a given iteration of the SIC process, an edge of a CN can be removed if and only if all remaining edges have been removed in previous iterations. Thus, CNs act as AND-nodes. Conversely, an edge of a VN can be removed if at least one of the remaining edges has been removed in previous iterations. Thus, VNs act as OR-nodes. Hence these trees are often referred to as AND-OR trees \cite{andor}.

AND-nodes are categorized into $s$ different types where each type corresponds to one of the subframes. Similarly, OR-nodes are categorized into $r$ different types where each type corresponds to one group of devices. For each graph $\mathcal{G}_s$, we consider $r$ trees where the $i^{th}$ tree is denoted by $T_{i,\ell}^{(s)}$ with depth $2\ell$ and a Type-$i$ OR-node at its root (depth 0). Initially, each Type-$i$ OR-node at depth $2\ell$ is assigned a value 0 with a probability $q_{i}^{(s)}[0]$ and is 1 otherwise. We are interested in finding the probability that the root of each tree evaluates to 0.
\subsection{ACK-All Transmission Scheme}
We define $\epsilon_{i}^{(s)}$ as the average fraction of devices in $\mathcal{C}_i$ that remain unresolved at the end of subframe $s$. The following lemma models the evolution of the average error probabilities in a given SIC iteration at the end of the transmission subframe $s$ for the ACK-All scheme (Fig. \ref{BP_compare}a) and calculates $\epsilon_{i}^{(s)}$ for $1\leq i,s\leq r$.
\begin{lemma}\label{lemma_all}
For the ACK-All scheme, the probability that a device from the group $\mathcal{C}_i$ remains unresolved at the end of subframe $s$ is given below as:
\begin{align*}
&\epsilon_{i}^{(s)}= \lim_{\ell\rightarrow \infty} q_{i}^{(s)}[\ell], \quad \text{where }\nonumber\\
&q_{i}^{(s)}[\ell] = \\
&\psi_{i}^{(s)}\left(1-\sum_{s'=1}^{s}\bar{c}_{i}^{(s')}\delta^{(s')}\left(1-\sum_{i'=1}^r{\bar{v}_{i'}^{(s')}\frac{q_{i'}^{(s)}[\ell-1]}{q_{i'}^{(s')}[0]}}\right)\right),\nonumber\\
&q_{i}^{(s)}[0] = \epsilon_{i}^{(s-1)}, \quad q_{i}^{(1)}[0] = 1,\nonumber\\
&\bar{v}_{i}^{(s)} = \frac{p_{i}^{(s)}|\mathcal{C}_{i,s}^{(s)}|}{\sum_{i'=1}^r p_{i'}^{(s)}|\mathcal{C}_{i',s}^{(s)}|} = \frac{g_i^{(s)}}{\sum_{i'=1}^r g_{i'}^{(s)}},\text{ and }\\
\end{align*}
\begin{align*}
&\bar{c}_{i}^{(s)} = \frac{p_{i}^{(s)}\Delta N_s}{\sum_{s'=1}^s p_{i}^{(s')}\Delta N_{s'}} = \frac{\zeta_i^{(s)}}{\sum_{s'=1}^s \zeta_i^{(s')}}.\\
\end{align*}
\end{lemma}
Readers are referred to Appendix \ref{app:1} for the proof of this lemma.
\subsection{ACK-Group Transmission Scheme}
In the case of separate transmissions, the transmissions of devices within a group are non-overlapping with the transmissions of devices from other groups. Therefore, the CNs in each subframe are only connected to one group of VNs. In other words, Type-$i$ AND-nodes are only connected to Type-$i$ OR-nodes, for $1\leq i \leq r$. Following on the proof of Lemma \ref{lemma_all}, the ACK-Group scheme is a special case of our proposed scheme where $g^{(s)}_i = 0$ for $i\neq s$. The following proposition is then derived.
\begin{proposition}\label{lemma_sep}
For the ACK-Group scheme, the probability that a device from the group $\mathcal{C}_i$ remains unresolved at the end of subframe $i$ is given below as:
\begin{align*}
&\epsilon_{i}^{(i)}= \lim_{\ell\rightarrow \infty} q_{i}[\ell], \text{ where }\nonumber\\
&q_{i}[\ell] =\lambda_{i}^{(i)}\left(1-\omega_i^{(i)}\left(1-q_{i}[\ell-1]\right)\right),q_{i}[0] = 1,\\
&\lambda_{i}^{(s)}(x) \triangleq \frac{1}{\sum_d d\Lambda_{i,d}^{(s)}}\sum_d d\Lambda_{i,d}^{(s)} x^{d-1} \text{ and }\\
 &\omega^{(s)}(x) \triangleq \frac{1}{\sum_d d\Omega_{i,d}^{(s)}}\sum_d d\Omega_{d}^{(s)} x^{d-1}.
\end{align*}
\end{proposition}
Another possible data transmission scheme is to acknowledge the resolved devices from $\mathcal{C}_i$ only at the end of subframe $i$. Based on the work in \cite{sorensenVTC} on codes-on-graph, authors showed that the encoding of already decoded information symbols is useful in the evolution of the decoding process at the receiver. However, we found that the transmissions of resolved devices in the following subframes are unnecessary and actually degrade the performance of the SIC process at the BS. The scheme performed poorly in comparison to the ACK-All scheme and demonstrated little performance gains in comparison to the ACK-Group scheme \cite{abbas2016performance}; therefore, we will not consider it in the paper.
\section{Design of Energy Efficient RMA Schemes}
\subsection{Performance Metrics}
For ease of notation, we denote the size of each group $\mathcal{C}_i$ by $\alpha_iK$ where $0<\alpha_i<1$ and $\sum_i\alpha_i = 1$. We also define a set of ratios $\beta_1,\beta_2,...,\beta_r$, where $\beta_i \triangleq \frac{N_i}{N}$. We now list the main performance metrics to be considered:
\subsubsection{System Load} The system load is denoted by $L$ and is defined as the ratio of the number of active devices $K$ to the number of time slots in the transmission frame $N$.
\subsubsection{Device Resolution Error} We define another QoS requirement, namely, the target average probability of device resolution error $\epsilon_i^*$, for $1\leq i \leq r$. This is the maximum acceptable fraction of devices from group $\mathcal{C}_i$ that violate their latency requirement, on average. That is, we need to ensure that $\epsilon_{i}^{(i)}\leq \epsilon_i^*$.
\subsubsection{Blocking Probability} In some cases the system load may be very large that there is no solution that can satisfy the QoS requirements, i.e., delay and probability of device resolution. The system is said to be overloaded. In this case, the BS enforces some access barring techniques \cite{survey1} to block some devices from accessing the network so as not to jeopardize the system performance. Details on this will be explained in Section \ref{acb_sec}.
\subsubsection{Average Number of Transmissions} M2M devices are notorious for having low power budgets. The average number of transmissions required for a device to meet its QoS requirement should be limited below a certain threshold for an energy efficient system. The average number of transmissions for each group can be calculated from the following proposition. Readers are referred to Appendix \ref{app:2} for the proof of this proposition.
\begin{proposition}\label{prop_Tr}
The average number of transmissions of a device from the group $\mathcal{C}_i$ can be calculated as
\begin{align*}
M_i=\sum_{s = 1}^r g_i^{(s)}\frac{\beta_s}{\alpha_i}\frac{N}{K}.
\end{align*}
\end{proposition}
\subsection{Design Objectives}
We denote by $\textbf{G}$ an $r\times r$ matrix given as
\begin{align*}
\textbf{G} = \begin{pmatrix}g_1^{(1)}&\ldots &g_r^{(1)}\\
                                     &\ddots&\\
                             g_1^{(r)}&\ldots&g_r^{(r)}\end{pmatrix}
\end{align*}
The main design objective is to find a matrix $\textbf{G}$ that satisfies the latency requirements with an acceptable average probability of device resolution error. In all what follows, for simplicity, we only consider strict latency requirements, i.e., $g_i^{(s)} = 0$ for $i>s$, which is a special case of the previously derived expressions. Therefore, we have to solve for $\sum_{i=1}^r r-i+1$ variables instead of $r^2$ variables. The main design objective becomes to find a matrix $\textbf{G}$ that satisfies the following constraint:
\begin{equation}\label{ct_r1}
\epsilon_i^{(i)} \leq \epsilon_i^*,\quad\text{ for } 1\leq i\leq r.
\end{equation}

Let $L^*(\epsilon)$ denote the maximum load for which a group of devices can be resolved within the required time with an average error probability of $\epsilon$. The load of each group in each subframe is defined as the ratio of the number of unresolved devices in that group to the number of time slots in that subframe. Based on this definition, we present the following proposition with the cases where it is possible to satisfy the constraints in (\ref{ct_r1}). Readers are referred to Appendix \ref{app:3} for proof of this proposition.
\begin{proposition}\label{prop_L}
There exists a matrix $\textbf{G}$ that can satisfy the latency requirements of all the groups in the ACK-All and ACK-Group schemes if and only if the following condition is satisfied:
\begin{align}\label{limit_load}
\frac{\epsilon_i^{(i-1)}\alpha_i}{\beta_i}\frac{K}{N}\leq L^*\left(\frac{\epsilon^*_i}{\epsilon_i^{(i-1)}}\right),
\end{align}
where $\epsilon_i^{(0)} = 1$ and $1\leq i \leq r$.
\end{proposition}
For the special case when the actual load of a particular group is approximately equal to the bound in (\ref{limit_load}), the devices of the respective group have to transmit separately to guarantee their QoS requirements in the two schemes: ACK-All and ACK-Group. Furthermore, when the actual load of all groups is approximately equal to $L^*$, the matrix $\textbf{G}$ that best satisfies the necessary constraints is the same for both schemes. In other words, the best performance is achieved with separate transmissions.

For the ACK-Group scheme, with reference to Proposition \ref{prop_L}, we have to ensure that $L^*(\epsilon_i^*)\geq\frac{\alpha_i K}{\beta_i N} $. Consider the special case when $\epsilon_i^* =\epsilon$ for $1\leq i\leq r$. We have the following design constraint
\begin{align}\label{ineq_sep}
\frac{K}{N}\leq L^*(\epsilon)\min\{\frac{\beta_1}{\alpha_1},\ldots,\frac{\beta_r}{\alpha_r}\}.
\end{align}
The arguments of the $\min$ function imply that some subframes may have a lower system load than others. Consider the case when the number of devices in group $\mathcal{C}_1$ is relatively small in comparison to its latency requirement. Then, it is likely that devices of the first group will be resolved with less than $\beta_1 N$ time slots. In fact, from the definition of $L^*(\epsilon)$, these devices can satisfy their QoS requirement with only $\frac{\alpha_1 K}{ L(\epsilon_1^*)}$ time slots. Therefore, for fair comparison, the length of every subframe $s<r$ is set to a length of $\min\left(\beta_s N,\frac{\alpha_s K}{L(\epsilon_s^*)}\right)$ which will allow the remaining groups more time slots to meet or improve their performance, if needed.
\subsection{Special case of r = 2}
\begin{figure}[!t]
  \centering
  \includegraphics[width=3.0in]{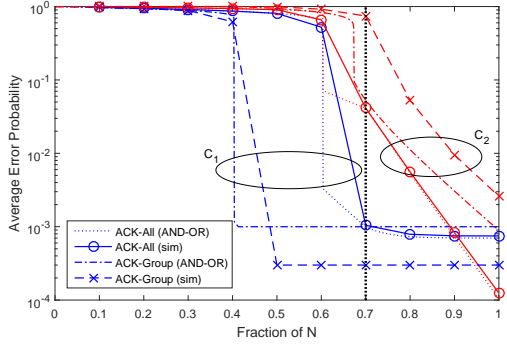}\\
  \caption{Achievable probabilities of device resolution error for two groups of MTC devices with $N/K = 2.0$, $\beta_1 = 0.7$, $\alpha_1 = 0.5$ and $N = 8000$.}
  \label{r2_load2}
\end{figure}
\begin{figure}[!t]
  \centering
  \includegraphics[width=3.0in]{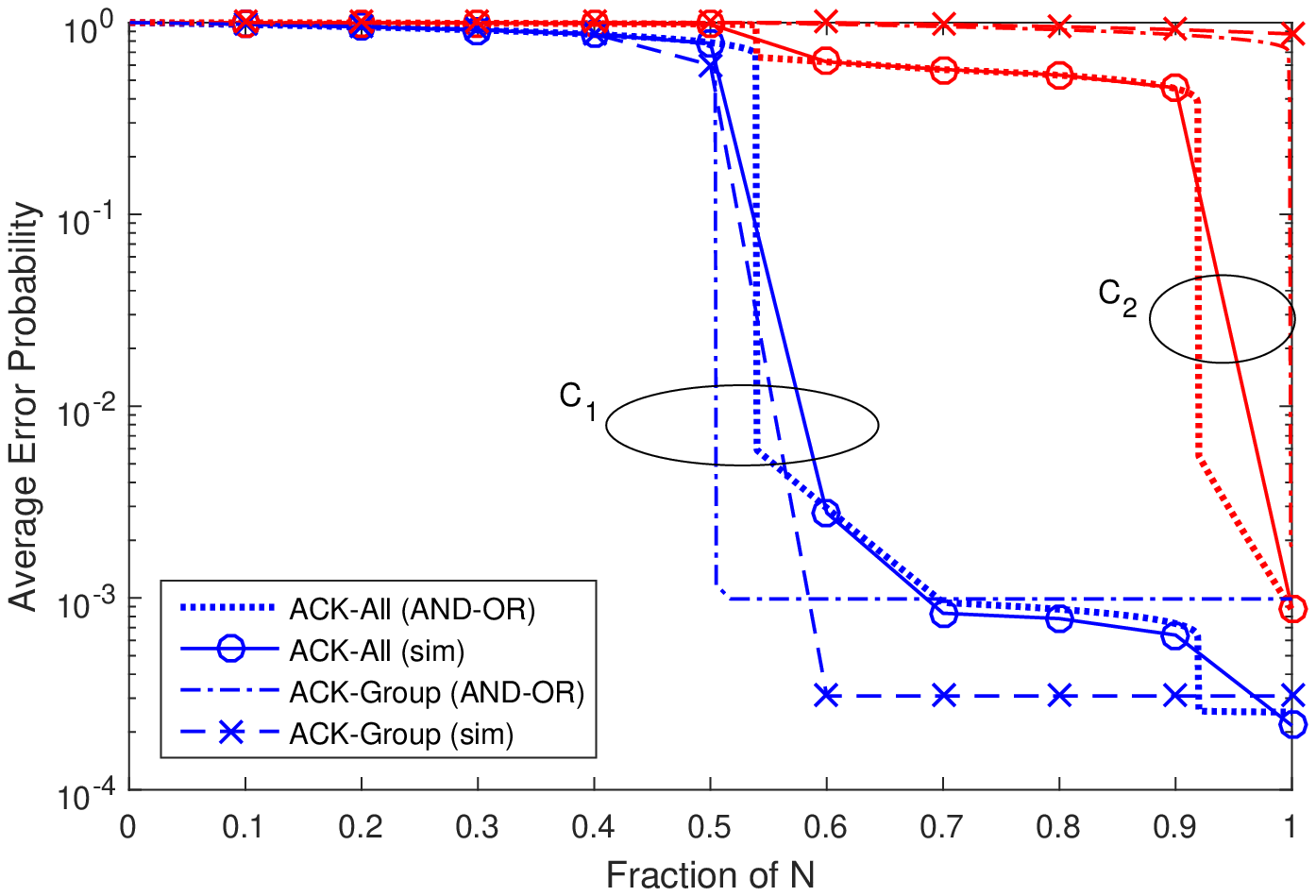}\\
  \caption{Achievable probabilities of device resolution error for two groups of MTC devices with $N/K = 1.6$, $\beta_1 = 0.7$, $\alpha_1 = 0.5$ and $N = 8000$.}
  \label{r2_load16}
\end{figure}
\begin{figure}[!t]
  \centering
  \includegraphics[width=3.0in]{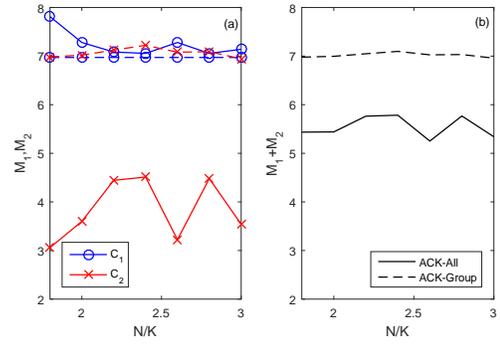}\\
  \caption{Average number of transmissions for two groups of MTC devices with $\beta_1 = 0.7$ and $\alpha_1 = 0.5$: (a) average number of transmissions of each group, (b) sum of average number of transmissions. The solid lines correspond to the ACK-All scheme, and the dashed lines correspond to the ACK-Group scheme.}
  \label{r2_tr}
\end{figure}
We compare the two transmission schemes for the same system loads and QoS requirements. Consider the case where $r=2$. We set $\epsilon^*_1 = \epsilon_2^* = 10^{-3}$ and use differential evolution \cite{de} to find $\textbf{G}$ that minimizes the sum of average transmissions $M_1 + M_2$ whose expressions were given in Proposition $\ref{prop_Tr}$. Fig. \ref{r2_load2} shows the analytical results for both schemes as well as the simulation results for $K = 4000$. The simulation results closely match those calculated from the AND-OR tree. For the ACK-All scheme, we can see that group $\mathcal{C}_1$ achieves the desired error probability of $10^{-3}$ after $\beta N$ time slots. As $\mathcal{C}_2$ participates in encoding in this subframe, $\mathcal{C}_1$ needs more time to reach its target error probability in comparison to the ACK-Group scheme where only devices of $\mathcal{C}_1$ participate in encoding. However, we see the disadvantage of ACK-Group in Fig. \ref{r2_load16}. The achievable error probabilities are plotted for a larger system load. In this figure, we can see that even though group $\mathcal{C}_1$ maintains the same performance, group $\mathcal{C}_2$ fails to satisfy its QoS with the ACK-Group scheme. Thus, we say that the ACK-All scheme can service more devices in this case. We will elaborate on this in Section \ref{design}.

{The other disadvantage of the ACK-Group scheme is shown in Fig. \ref{r2_tr}(a). Although $M_1$ is the same in both schemes, $M_2$ is reduced in the ACK-All scheme leading to a reduction in the overall average number of transmissions of all devices in the system as in shown in Fig. \ref{r2_tr}(b). As the average number of transmissions is related to the energy expenditure, we say that the ACK-All scheme is more energy efficient in this case. Nevertheless, we can see that the average number of transmissions for both RMA schemes is quite comparable to the number of transmissions allowed in cellular access networks over the RACH \cite{shirvanimoghaddam2015probabilistic} which further supports RMA as a good candidate for future M2M communication networks.}
\subsection{Access Barring}\label{acb_sec}
Finally, let $L_r^*(\boldsymbol{\alpha},\boldsymbol{\beta},\boldsymbol{\epsilon^*})$ be the maximum system load that the BS can service for the parameters $\boldsymbol{\alpha} = [\alpha_1,\alpha_2,...,\alpha_r]$, $\boldsymbol{\beta} = [\beta_1,\beta_2,...,1]$ and constraints $\boldsymbol{\epsilon^*} = [\epsilon^*_1,\epsilon_2^*,...,\epsilon_r^*]$. Conventionally, whenever a device has a packet to transmit, it needs to locate and synchronize to a suitable BS based on its broadcast information. When the number of active devices is larger than $L_r^*N$, granting all devices access into the network will jeopardize the performance of all groups. In this case, the system may implement some form of access barring to limit the number of active devices in a given transmission frame. A simple example is given in the proposition below.
\begin{figure}
  \centering
  \includegraphics[width=3.0in]{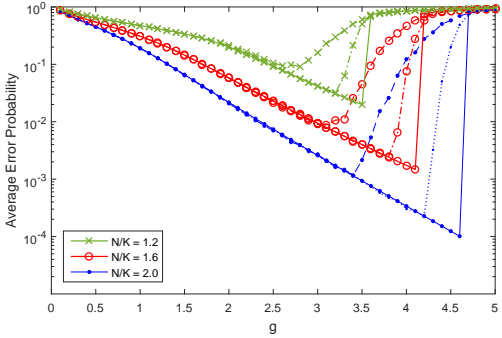}\\
  \caption{{Average error probabilities achievable by different access probabilities and different loads. The solid lines represent the error probabilities calculated from the AND-OR tree expressions. The dashed lines and the dash dotted lines represent the error probabilities obtained from simulations for $N = 200$ and $N = 2000$, respectively.}}\label{set1}
\end{figure}
\begin{proposition}\label{prop_acb}
Let $\frac{N}{K}$ be the system load at the beginning of a transmission frame. These devices are blocked with a probability of $b$ where $b$ is calculated below as
\begin{equation}
b = 1- \min\left(1,L_r^*(\boldsymbol{\alpha},\boldsymbol{\beta},\boldsymbol{\epsilon^*})\frac{N}{K}\right).
\end{equation}
\end{proposition}
\noindent Given $\boldsymbol{\alpha},\boldsymbol{\beta},\boldsymbol{\epsilon^*}$ and $N$, the system can guarantee the required QoS requirements for at most $L_r^*(\boldsymbol{\alpha},\boldsymbol{\beta},\boldsymbol{\epsilon^*})N$ devices. Proposition \ref{prop_acb} shows that when $K\leq L_r^*(\boldsymbol{\alpha},\boldsymbol{\beta},\boldsymbol{\epsilon^*})N$, all active devices are allowed access into the system ($b=0$). On the other hand, when $K>L_r^*(\boldsymbol{\alpha},\boldsymbol{\beta},\boldsymbol{\epsilon^*})N$, an average of only $(1-b)K = L_r^*(\boldsymbol{\alpha},\boldsymbol{\beta},\boldsymbol{\epsilon^*})N$ active devices are allowed access into the system. Otherwise, the system will be overloaded and will fail to satisfy the QoS requirements.  This access barring technique is often referred to as dynamic access barring (DAB) as the probability is updated in each transmission frame based on the current load and is not fixed a priori. It is worthy of noting that access class barring (ACB) schemes \cite{survey1} can also be implemented to block more devices of the less important applications rather than all applications equally.
\section{Design of Reliable RMA Schemes for a Finite Number of Devices}\label{design}
When concerned with maximizing the system reliability, i.e., minimizing the device resolution error probabilities, the inaccuracy of the asymptotic AND-OR tree results for finite values of $K$ and $N$ has been noted in previous works \cite{popovskiLetter}. In this section, we elaborate on this discrepancy and propose a guideline that enables us to use the AND-OR tree expressions to design the access probabilities for a finite number of devices. We start off by considering only a single group ($r=1$) of $K$ devices transmitting with an access probability of $\frac{g}{K}$, where $g=g_1^{(1)}>0$ is the average number of devices that access a given time slot.

Let us denote by $\epsilon\left(g,\frac{K}{N}\right)$ the average probability of device resolution error $\epsilon_1^{(1)}$ (ref. Section II) when the system load is $\frac{K}{N}$ and the access probability is $\frac{g}{K}$. Here, we add the arguments $g$ and $\frac{K}{N}$ to distinguish between the error probabilities achievable under different system loads and using different access probabilities. It is also to emphasize that they are the only parameters necessary to calculate the average probability of device resolution error in the asymptotic case.
{We calculate $\epsilon\left(g,\frac{K}{N}\right)$ from Proposition \ref{lemma_sep} under different settings and plot the results in Fig. \ref{set1} along with the simulation results for $N=200$ and $N= 2000$. Let us denote by $g^*$ the corresponding values of $g$ at the extrema points, i.e., $g^* = \arg {\min_{g} \epsilon\left(g,\frac{K}{N}\right)}$. We notice that for both values of $N$ there is a significant mismatch between the simulation results and the AND-OR expressions in the region $[g^*-\sigma, g^*+\sigma]$, where $\sigma$ is a positive decreasing function of $N$, i.e., $\sigma \rightarrow 0$ when $N\rightarrow \infty$.}
\begin{figure}
  \centering
  \includegraphics[width=3.0in]{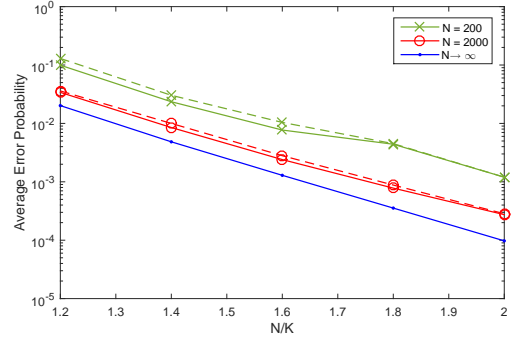}\\
  \caption{{Average error probabilities achievable by the proposed design guideline for $c = 10$. The solid lines represent the minimum error probabilities achievable in simulations and the dashed lines represent the error probabilities achieved by the design.}}\label{des_res}
\end{figure}

{Moreover, we also observe that the error probabilities for all loads decrease gradually as $g$ approaches the extrema points $g^*$. On the other hand, the error probability curves beyond that point become sharply increasing. For example, at $N/K = 1.2$, the error probability jumps from $0.02$ to $0.61$ when $g$ increases from $3.49$ to only $3.50$. That is, in that region, the achievable error probability is very sensitive to the variations in $g$. Now, recall from Fig. \ref{BP2}a, that the probability of an edge connecting a pair of VN and CN is Bernoulli distributed with a success probability of $\frac{g}{K}$. Therefore, the sum of edges in a bipartite graph is a Poisson distributed random variable with an average of $gN$. Accordingly, we can generate an infinite number of random bipartite graphs given the parameters $K$, $N$, and $g$. Let us denote by $X$ the sum of edges of a random bipartite graph, and let $G= \frac{X}{N}$. Then, $\frac{G}{K}$ denotes the effective access probability of this bipartite graph. It is straightforward to see that $G$ is also a Poisson distributed random variable with an average of $g/N$. Thus, the variations in $G$ and the effective access probability of the graph increases with the decrease in $N$. For example, for $g\leq 4$, the standard deviation of $G$ would be approximately $0.14$ and $0.04$ for $N = 200$ and $N = 2000$, respectively. Therefore, a system with a finite number of devices operating at the optimal points $g^*$ will exhibit large variations in system performance leading to the discrepancies observed between the actual average system performances and that predicted by the AND-OR tree in Fig. \ref{set1}.}
\begin{figure}[!t]
  \centering
  \includegraphics[width=3.0in]{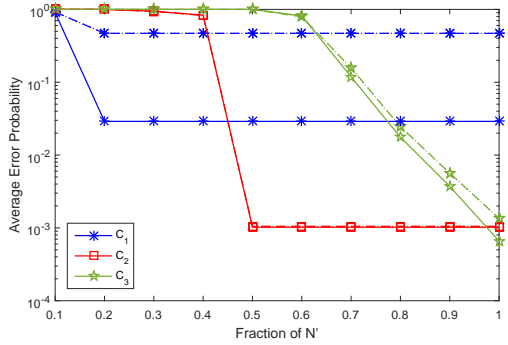}\\
  \caption{Achievable probabilities of device resolution error by each scheme for $N'/K = 2$, $\boldsymbol{\beta} = [0.2, 0.5, 1]$ and $\boldsymbol{\alpha} = [\frac{1}{3}, \frac{1}{3}, \frac{1}{3}]$. The solid lines and dotted lines correspond to the ACK-All scheme  with c = 10 and c = 0, respectively. The dashed lines correspond to the ACK-Group scheme  with c = 10.}\label{equal}
\end{figure}

{Based on all the above, it is now clear why the optimal values of $g^*$ for a finite number of devices cannot be calculated directly from the AND-OR tree expressions. This behavior was noted in \cite{popovskiLetter} and was avoided by conducting an exhaustive search of the optimal access probabilities. While this approach may be acceptable for a small number of devices, it becomes very tedious with the increase in the number of devices as well as the number of groups. We find that a more accurate calculation of the achievable average error probability at $g$
for a given number of devices $K$ and number of time slots $N$ is possible by taking the average of all three values: $\epsilon\left(g-\sigma,\frac{K}{N}\right)$, $\epsilon\left(g,\frac{K}{N}\right)$ and $\epsilon\left(g+\sigma,\frac{K}{N}\right)$, where $\sigma = c\sqrt{\frac{g}{N}}$ for some positive constant $c$. The accuracy of our proposed design is shown in Fig. \ref{des_res}.
We make note that was not necessary in the previous section as we were concerned with minimizing the number of transmissions rather than the error probabilities. In other words, we were not concerned with operating at the extrema points.}

For the sake of completion, we extend our design to the case of multiple QoS requirements with 3 groups of devices. We set the acceptable average probability of device resolution error to $10^{-3}$ for all groups. Furthermore, for the sake of energy efficiency, we limit the average number of transmissions, we only consider ranges of $0\leq g_i^{(s)}\leq 4$. In Fig. \ref{equal}, we consider the case where all the groups in the system contain the same number of devices. We plot the achievable error probabilities with access probabilities designed directly from the AND-OR tree expressions and those designed using the aforementioned guideline. First of all, we observe the biggest mismatch between the two approaches for $\mathcal{C}_1$ whose minimum average probability of device resolution error is above the required threshold. This validates the importance of our proposed guideline in determining suitable points of operation.

\begin{figure}[!t]
  \centering
  \includegraphics[width=3.0in]{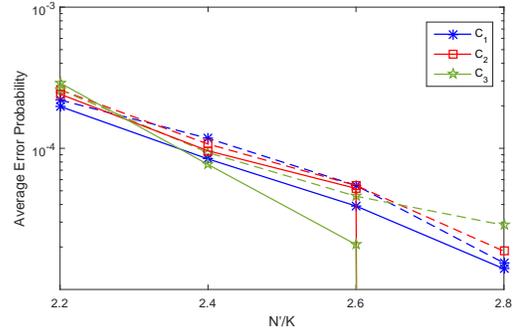}\\
  \caption{Achievable probabilities of device resolution error by each scheme for $\boldsymbol{\alpha} = [0.2, 0.3, 0.5]$. The solid lines correspond to the ACK-All scheme, and the dashed lines correspond to the ACK-Group scheme with c = 10.}\label{prop}
\end{figure}

Interestingly, we observe that both schemes demonstrate the same performance. This validates the results of Proposition \ref{prop_L} that non-separate transmissions are only possible when the load per each group is larger than the given bound. Otherwise, the optimal performance converges to that of the ACK-Group scheme. In Fig. \ref{prop}, we consider the same setup but for the case where the number of devices in each group is almost proportional to its delay requirement. In this case, the load per group is significantly larger than the bound in (\ref{limit_load}) which allows for the sharing of resources. We plot the achievable performance of each group for different loads. For $N/K$ approximately larger than 2, the ACK-All scheme can significantly improve the performance of $\mathcal{C}_3$ while guaranteeing $\mathcal{C}_1$ and $\mathcal{C}_2$ their required thresholds. Such an assumption on the sizes of the groups is practical when considering a Poisson packet arrival model where all devices arrive at the same rate. As devices with tighter delay requirements are served faster than others, the number of queuing devices in $\mathcal{C}_i$ at the beginning of each transmission frame will be on average always less than those in $\mathcal{C}_j$, for $i<j$.
\section{Performance Evaluation}
\subsection{LTE Setting}
{In this section, we evaluate the performance of RMA in an LTE-based setting \cite{wiriaatmadja2015hybrid}. In each transmission frame, a number of resource blocks (RBs) is allocated for M2M communications. In LTE, an RB is the smallest radio resource unit that can be allocated to a device. It is made up of one time slot and one subchannel. Thus, the incorporation of RMA into an LTE-based setting requires an efficient resource management scheme of the two-dimensional resources. Although this is outside the scope of our work, we show that even with a simple setup, RMA can still score better performance over coordinated access.}

{As  access barring is said to be the best currently available solution to the RACH overload problem and has been used in the standardization process of LTE-A \cite{cheng2011prioritized}, we consider two setups for access barring as benchmarks for comparison. In the first setup, the number of RBs allocated to the RACH in each frame is fixed a priori. In the second setup, the fraction of RBs allocated to the RACH in each frame is fixed a priori. However, for both setups, we consider DAB where the blocking probability is updated in each frame based on the current load. Devices that are blocked or unsuccessfully resolved are allowed to retransmit in the following frame. It was shown before that access barring and even DAB will not suffice as a stand alone solution in future cellular networks \cite{survey2}; therefore, we consider the work of \cite{wiriaatmadja2015hybrid} which combines DAB with the dynamic resource allocation of RACHs and data channels as another benchmark. In our simulation setup, we assume that we can construct 8 preambles from each RB and that each packet can be transmitted within one RB.}

\begin{figure}[!t]
\centering
  \centering
  \includegraphics[width=3.0in]{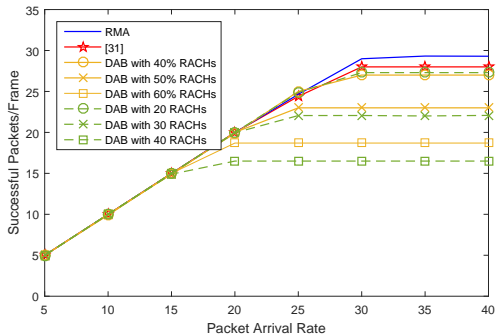}
  \caption{Number of successfully recovered packets per frame for RMA, DAB and the dynamic resource allocation scheme in \cite{wiriaatmadja2015hybrid}}\label{throughput1}
\end{figure}

{For RMA, we use $N$ to denote the number of RBs in a frame which is not necessarily the number of time slots. In fact, each transmission frame is assumed to be fixed in time duration, and the packet arrival rate $\lambda$ is defined as the number of packets per frame. Fig. \ref{throughput1} shows the throughput of the system defined as the average number of resolved devices per transmission frame. In each transmission frame, the number of RBs reserved for M2M communications is a uniformly distributed random variable varying from 0 to 100. Simulation results are plotted for different packet arrival rates. The average number of resolved devices is shown to increase for all schemes with the increase in $\lambda$ until it reaches a certain saturation point. We observe that RMA can achieve a higher throughput and, thus, can support a larger packet arrival rate. }

\begin{figure}[!t]
  \centering
  \includegraphics[width=3.0in]{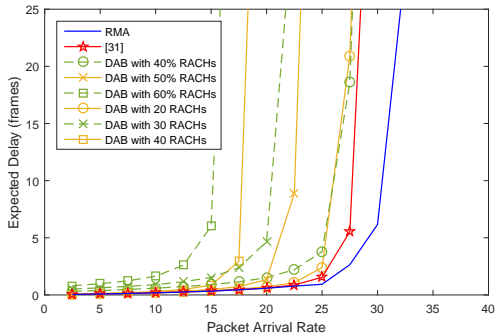}
  \caption{Expected delay in number of frames for RMA, DAB and the dynamic resource allocation scheme in \cite{wiriaatmadja2015hybrid}} \label{delay1}
\end{figure}

\begin{figure}[!t]
  \centering
  \includegraphics[width=3.0in]{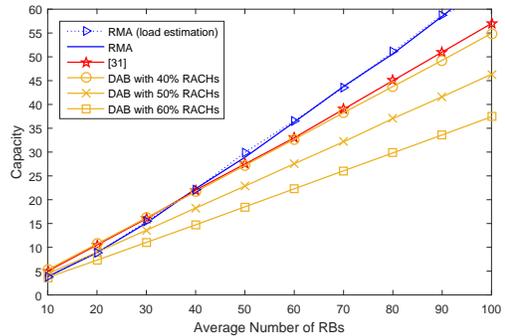}
  \caption{Maximum achievable throughput for different average number of RBs.}\label{capacity1}
\end{figure}

{When the throughput saturates, the system  is said to be unable to service all active devices and the blocking probability will start to increase. As more and more devices are barred from access, the expected delays increase, and the system becomes unstable. We note that the system is said to be stable provided that the expected delay experienced by the devices is bounded below a certain threshold. This is shown in Fig. \ref{delay1}, where the expected delay is expressed as the number of transmission frames. We can see that again RMA can support a larger packet arrival rate while guaranteeing small latencies. Finally, in Fig. \ref{capacity1}, we show the capacity of these schemes for different average number of RBs. The capacity is defined as the maximum throughput that can be stably supported. The capacity of RMA is shown to be the largest when the number of resources available is sufficiently large.}

\subsection{Practical Considerations}
We now shed light on a number of important practical considerations that arise with the considered framework.
\subsubsection{Load Estimation}
{For batch arrivals, there are a number of existing load estimation algorithms based on statistical inferences that can be incorporated into this work, e.g., \cite{FASA}. Similar inferences can be used for the case of continuous arrivals. For example, we can incorporate the number of resolved devices in each frame into the estimation. Let $K[i]$, $K_s[i]$ and $b[i]$ be the estimated number of participating devices, the number of successfully resolved devices and the access barring probability, in frame $i$, respectively. Then, the estimated number of participating devices can be expressed as $K[i] = \lambda + \left((1-b[i-1])K[i-1]-K_s[i-1]\right) + b[i-1]K[i-1]$, where the first, second and third terms correspond to the new packet arrivals in frame $i$, the unsuccessfully recovered packets in frame $i-1$ and the barred packets in frame $i-1$, respectively. In Fig. \ref{capacity1}, we show that the performance degradation due to inaccurate load estimation is quite negligible. This estimate can be further improved by considering the statistical information of the previous singleton, collision, and idol slots. Performance can also be improved by designing the system for a value of $K(1+\rho)$, where $\rho$ is the fractional offset error in estimation.}
\subsubsection{Lossy Feedback Channel}
\begin{figure}[!t]
\centering
  \centering
  \includegraphics[width=3.0in]{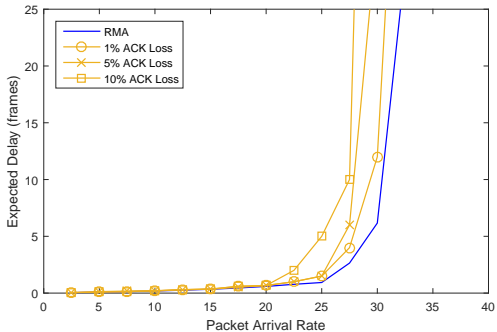}
\caption{The effect of lossy feedback channels on the stability of RMA}\label{prac}
\end{figure}

{So far, we have assumed a perfect feedback. However, in some cases, acknowledgements to successfully resolved devices might be lost in the network. These unacknowledged devices will retransmit in the following subframes/frames assuming their packets have not been successfully recovered yet. This will induce a dynamic behavior in the system and can affect the stability of the system.} We investigate the impact of lossy feedback channels on the system performance in Fig. \ref{prac}. We can see that with an imperfect feedback channel,
{the losses in performance are minimal provided that the losses in acknowledgements are relatively small ($<0.01$), as is generally assumed. Moreover, knowing the feedback channel state, one may redesign the system to a target error probability equal to the product of that of the SIC and that of the feedback channel.}\\
{We refer the readers to the work in \cite{popovskiARXIV} for a study on the reliability of control information and techniques to increase this reliability. In addition, we emphasize that some of the newly proposed paradigms are considering open-loop communications, where packets are not acknowledged. In this case, device transmissions are either limited in number or within a certain time frame, after which the devices either start transmitting another packet or become idle. Such settings have been shown to have significant gains in terms of network latency in comparison to closed-loop communications \cite{H-CRAN}}
\subsubsection{Channel Estimation}
{Channel estimation is necessary in RMA for two main reasons. First of all, based on the received power and known CSI, the BS can distinguish between idle slots, singleton slots and collision slots. Second of all, the the received power needs to be large enough to allow correct decoding of the information for the given coding and modulation scheme at the receiver. Therefore, our scheme does not require perfect channel estimation.}\\
{One should note that coded random access only attempts to recover a packet when it is in a singleton slot. As mentioned before, the detection of a singleton slot is based solely on the received power level. Therefore, issues with inter-symbol interference can be resolved using the same techniques used for point to point scenarios. We assume the inter-symbol interference caused by the overlapping between different packets in different slots will have little impact on the SIC performance due to the very low interference levels for small synchronization errors. To prevent such misalignment, some guard zone can be added at the beginning and the end of each symbol. Since we assume small synchronization errors, the duration of the guard zone in each symbol is also negligible. Furthermore, once a packet is recovered from a singleton slot, the BS can determine its timing offset. As the devices are assumed to be static for the duration of the transmission frame, the BS can use this information to accurately cancel the packet out from other collision slots.}
\section{Conclusion}
In this paper, we proposed a random multiple access scheme with QoS guarantees for a heterogenous M2M communication network. We considered two transmission schemes: ACK-All and ACK-Group. The ACK-All scheme allows for the simultaneous transmissions of different device groups over the same resources. Meanwhile, the ACK-Group scheme assumes that devices from different groups transmit over different resources. We drew an analogy between our proposed scheme and the codes on graph, and we derived the expressions for the average probability of device resolution for each of these two schemes based on the AND-OR tree. We showed the accuracy of these expressions in calculating the error probabilities and proposed a guideline to design the access probabilities in practical M2M settings. We showed that non-separate transmissions are only beneficial when the loads of the groups with tighter deadlines is relatively low to enable them to share their resources with the remaining groups. Otherwise, we showed through analysis and simulations that these schemes yield the same performance as the ACK-Group scheme. Finally, we showed that the proposed scheme is superior to coordinated access schemes when the number of active devices and available resources is large enough.
\appendices
\section{Proof of Lemma 1}\label{app:1}
Consider an arbitrary graph $\mathcal{G}_s$ constructed with the ACK-All scheme. Let $q_{i}^{(s)}[\ell]$ be the probability that a Type-$i$ OR-node remains unresolved in the $\ell^{th}$ iteration of the SIC process. This probability is initialized as $q_{i}^{(s)}[0] = \frac{|\mathcal{C}_{i,s}^{(s)}|}{|\mathcal{C}_i|}$. AND-nodes are divided into $s$ types corresponding to the $s$ subframes. A Type-$i$ AND-node is said to have $d$ children with a probability $\delta_{d}^{(i)}$. On the other hand, OR-nodes are divided into $r$ types corresponding to the $r$ groups. Each type of OR-node is further divided into $s$ sub-Types corresponding to the $s$ subsets, where the first $s-1$ sub-Types have a value of 1 (having been resolved previously). A Type-$i$ OR-node is said to have $d$ children with a probability of $\psi_{i,d}^{(j)}$.

A Type-$i$ OR-node is said to be connected to a Type-$j$ AND-node with a probability $\bar{c}_i^{(j)}$. This probability is given below as the normalized selection probability of the corresponding subframe by group  $i$.
\begin{equation*}
\bar{c}_{i}^{(j)} = \begin{cases}\frac{p_{i}^{(j)}\Delta N_j}{\sum_{j'=1}^j p_{i}^{(j')}\Delta N_{j'}} = \frac{\zeta_i^{(j)}}{\sum_{j'=1}^j \zeta_i^{(j')}}&\text{for sub-Types } 1,...,j\\0&\text{otherwise}\end{cases}.
\end{equation*}
From Fig. \ref{BP2}, we find that the degrees of some AND-nodes are reduced in each subframe as more of their edges are removed. Based on (\ref{CN_bino}), the probability that a Type-$j$ AND-node is still connected to $d$ Type-$i$ OR-nodes in $\mathcal{G}_s$ is given as :
\begin{align*}\label{CN_red}
\Omega_{i,d}^{(j\rightarrow s)} &= \begin{pmatrix} |\mathcal{C}_{i,s}^{(s)}|\\d\end{pmatrix}\left(\frac{g_i^{(j)}}{|\mathcal{C}_{i,j}^{(j)}|}\right)^d\left(1-\frac{g_i^{(j)}}{|\mathcal{C}_{i,j}^{(j)}|}\right)^{|\mathcal{C}_{i,s}^{(s)}|-d}.
\end{align*}
For $K,N \gg$, the Poisson approximation of the former equation is given as
\begin{equation}\label{CN_poiss_red}
\Delta^{(j\rightarrow s)}(x) = \exp\left(-\sum_{i=1}^{r}\frac{q_i^{(s)}[0]}{q_i^{(j)}[0]}g_{i}^{(j)}(1-x)\right),
\end{equation}
Moreover, given that an edge of a Type-$j$ AND-node has not been removed in the previous $\ell-1$ iterations of the SIC, this edge is said to be connected to a Type-$i$ OR-node with a probability $\bar{v}_{i}^{(j\rightarrow s)}[\ell]$. $\bar{v}_{i}^{(j\rightarrow s)}[\ell]$ is expressed below as the normalized access probability of the corresponding subset in subframe $j$.
\begin{equation}
\bar{v}_{i}^{(j\rightarrow s)} = \frac{p_i^{(j)}|\mathcal{C}_{i,s}^{(s)}|}{\sum_{i'=1}^s p_{i'}^{(j)}|\mathcal{C}_{i',s}^{(s)}|}= \frac{\frac{q_i^{(s)}[0]}{q_i^{(j)}[0]}g_i^{(j)}}{\sum_{i'=1}^s \frac{q_{i'}^{(s)}[0]}{q_{i'}^{(j)}[0]}g_{i'}^{(j)}}.
\end{equation}

In general, each AND-Node at depth $2\ell-1$ calculates its value by performing AND operation on its children at depth $2\ell$. Assume that a child of an AND-node has a value of $1$ with a probability $1-q_c$. Then, the probability that an AND-node of degree $d$ has a value of $0$ is $1-(1-q_c)^d$. Similarly, each OR-node at depth $2\ell$ calculates its value by performing OR operation on its children at depth $2\ell+1$. Assume that a child of an OR-node has a value of $1$ with a probability $q_v$. Then, the probability that an OR-node of degree $d$ has a value of $0$ is $(q_v)^d$. More generally, we can write
\begin{align*}
1-q_{c,j} &= \sum_{i=1}^r\bar{v}_{i}^{(j\rightarrow s)}\frac{q_{v,i}}{q_i^{(s)}[0]},
\end{align*}
where $q_{c,j}$ is the probability that a child of an AND-node of Type-$j$ has a value of 0. Similarly, $q_{v,i}$ is the probability that a child of an OR-node of Type-$i$ has a value of 1.
By averaging this expression over the degree distribution in (\ref{CN_poiss_red}) and the different types of AND-nodes, we arrive at
\begin{align*}
q_{v,i}  &= \sum_{j=1}^s \bar{c}_i^{(j)}\sum_d\delta^{(j\rightarrow s)}_d\left(1-\left(1-q_{c,j}\right)^d\right)\\
&= 1-\sum_{j=1}^s \bar{c}_i^{(j)}\sum_d\delta^{(j\rightarrow s)}_d\left(1-q_{c,j}\right)^d\\
&= 1-\sum_{j=1}^s \bar{c}_i^{(j)}\delta^{(j\rightarrow s)}\left(1-q_{c,j}\right).
\end{align*}
Similarly, by averaging $q_{v,i}$ over the degree distribution in (\ref{VN_poiss_all}), we arrive at
\begin{align*}
q_{v,i}  = \sum_d\psi_{i,d}^{(s)}\left(q_{v,i}\right)^d = \psi_i^{(s)}(q_{v,i})
\end{align*}
Finally, with a slight modification in notations, it is straightforward to arrive at the expression in Lemma 1.
\section{Proof of Proposition \ref{prop_Tr}}\label{app:2}
In the ACK-All scheme, devices from the group $\mathcal{C}_i$ continue to transmit in following subframes provided they have not been resolved in previous subframes. Let $a_{i,n}^{(s)}$ be a Bernoulli random variable denoting the probability that a device $D$ from the group $\mathcal{C}_i$ has transmitted in the $n^{th}$ time slot of the $s^{th}$ subframe. Then, the average number of transmissions of a device from the group $\mathcal{C}_i$, denoted by $M_i$, is derived as the sum of average number of transmissions of this device in each subframe.
\begin{align*}
M_i &= \mathbb{E}\left[\sum_{s=1}^r\sum_{n=1}^{\Delta N_s}a_{i,n}^{(s)}\right] = \sum_{s=1}^r\mathbb{E}\left[\sum_{n=1}^{\Delta N_s}a_{i,n}^{(s)}\right]\\
&=\sum_{s=1}^r \Delta N_s \text{Pr}\left(\left.D\in\mathcal{C}_{i,s}^{(s)}\right|D\in\mathcal{C}_i\right)\text{Pr}\left(\left.a_{i,n}^{(s)}=1\right|D\in\mathcal{C}_{i,s}^{(s)}\right)\\
&=\sum_{s = 1}^r\Delta N_s\frac{g_i^{(s)}}{|\mathcal{C}_{i,s}^{(s)}|}\frac{\left|C_{i,s}^{(s)}\right|}{|\mathcal{C}_{i}|}
=\sum_{s = 1}^r\Delta N_s\frac{g_i^{(s)}}{|\mathcal{C}_{i}|}.
\end{align*}
\section{Proof of Proposition \ref{prop_L}}\label{app:3}
Following on the analogy between our RMA schemes and codes on graph, the system load is analogous to the code rate. The code rate is the ratio of the number of information symbols to the number of coded symbols. The channel capacity dictates the maximum achievable code rate for a given channel SNR at which reliable transmission is possible. Similarly, the bound $L^*(\epsilon)$ dictates the maximum system load for a given target error probability $\epsilon$ at which a reliable communication is possible. While the channel capacity can be calculated from the Shannon bound, $L^*(\epsilon)$ can be calculated from the AND-OR tree equations given in Proposition \ref{lemma_sep}. For example, in Fig. \ref{set1}, the achievable error probabilities are shown for different loads and different values of $g$ for the case of $r=1$. For $\epsilon = 0.02$, the system load is said to be upper bounded by $1/1.2$, i.e., there exists a $g>0$ that can guarantee an average probability of device resolution error of 0.02 for every $K/N$ less than $1/1.2$. It is straightforward to generalize this concept for $r>1$ for the ACK-Group scheme. Moreover, as the ACK-Group scheme is a special case of the ACK-All scheme, we can say that if there exists a matrix $\textbf{G}$ that can satisfy the latency requirements of all the groups in the ACK-Group scheme, then, there exists a matrix $\textbf{G}$ that can satisfy these requirements in the ACK-All scheme as well. The argument of $L^*$ in (\ref{limit_load}) takes into consideration that some of the devices of $\mathcal{C}_i$ have been resolved in the previous $i-1$ subframes. Therefore, the error probability is scaled according to the number of unresolved devices. For the ACK-Group scheme, $\epsilon_i^{(s)} = 1$ for $i\leq s$.

\bibliographystyle{IEEEtran}
\bibliography{M2M}
\end{document}